\documentclass[twocolumn,preprintnumbers,amsmath,amssymb,aps,prl,superscriptaddress,]{revtex4}
\pdfpageattr{/Group <</S /Transparency /I true /CS /DeviceRGB>>}

\usepackage{amssymb,amsfonts,amsmath}
\usepackage{euscript}
\usepackage{graphics}
\usepackage{setspace}
\usepackage{graphicx}
\usepackage{hyperref}
\usepackage{setspace}
\usepackage{dcolumn}
\usepackage{bm}
\usepackage{color}

\newcommand{\bT}{{\bf T}}

\begin{document}

\title{Periodic and chaotic orbits of plane-confined micro-rotors in creeping flows}

\author{Enkeleida Lushi}
\email{enkeleida\_lushi@brown.edu}
\author{Petia M. Vlahovska}
\affiliation{School of Engineering, Brown University, Providence, RI, 02912, USA}

\begin{abstract}

We explore theoretically the complex dynamics and emergent behaviors of spinning spheres immersed in viscous fluid. The  particles are coupled to each-other via the fluid in which they are suspended: each particle disturbs the surrounding fluid with a rotlet field and that fluid flow affects the motion of the other particles. We notice the emergence of intricate periodic or chaotic trajectories that depend on the rotors initial position and separation. The point-rotor motions confined to a plane bear similarities the classic 2D point-vortex dynamics. Our analyses highlight the complexity of the interaction between just a few rotors and suggest richer behavior in denser populations. We discuss how the model gives insight into more complex systems and suggest possible extensions for future theoretical studies.

\end{abstract}


\maketitle

Collective motion of active particles has received considerable recent interest in the scientific community, spanning across many disciplines and studied using diverse tools ranging from theoretical and computational to experiments. {\it Active matter} is composed of driven units or particles, each capable of converting stored or ambient energy into movement \cite{Marchetti13}. The interaction of such self-driven units with each-other and the surrounding environment gives rise to intriguing spontaneous organizations in their population. Biological active systems include mixtures of biofilaments and motor proteins from cell extracts, the cytoskeleton of living cells, bacterial suspensions, herds of terrestrial animals such as ants, aquatic animals such as fish, and aerial flocks of birds . Nonliving active matter includes vibrated granular rods, an ensemble of robots, suspensions of colloids propelled in fluid by their surface catalytic activity when chemicals are added to the system, or colloids driven into motion by light. For more information on these examples of active systems see the recent reviews on active and driven matter \cite{Marchetti13, Menzel15} and the references therein.

In the aforementioned active systems the individual units are self-propellers and in most instances the generated propulsion is translational.  Little attention has been shown on active units that rotate due to internal or external torques, partly because such types of rotating units were realized experimentally only recently. New experimental realizations of spinning particles have been reported when the particle rotation is driven by chemicals \cite{Mallouk09}, optical tweezers or light \cite{Grier97, Friese98}, magnetic \cite{Grzybowski00, Grzybowski02} or electric \cite{Bricard13} fields.  Increased interest in rotor systems generated theoretical studies exploring simple sphere rotors pair dynamics \cite{Leoni10, Fily12b},  self-assembly \cite{Climent07},  dynamics at interfaces \cite{Llopis08}, rheology of suspensions \cite{Yeo10b, Jibuti12}, chiral motor suspensions \cite{Fuerthauer13} or synchronization in a carpet of hydrodynamically coupled rotors with random intrinsic frequencies \cite{Najafi10}.

Other than direct particle collisions, the interactions in such systems are dominated by the fluid flow, i.e., at leading order a rotor particle is advected by the flow generated by the other rotors. The fundamental interactions between such particles and the resulting collective  dynamics  has  not been investigated much, with the notable exception of a pair of rotors \cite{Leoni10, Fily12b, Das13}. The pair dynamics has been studied on a more complex model of an active rotor, for example ones with an elongated shape \cite{Leoni10}, rotors which may be translating while also rotating \cite{Fily12b} or Quincke rotor pairs with electro-hydrodynamical interactions \cite{Das13}. The collective dynamics has been studied in more complex flow conditions, e.g. shear \cite{Yeo10b}. Surprisingly, the self-organization and coupled behavior of many rotor particles under no imposed external flow (i.e. in quiescent flow) has not been investigated. 

We study theoretically the collective dynamics of rotating particles and the fundamental interactions which lead to group behavior. For rotating particles suspended in viscous fluid the rotlet flows dominate the interactions. We show how the motion of two and three particles results in regular or periodic motion, as do certain symmetrical configurations of more particles. For four or more rotlets, irregular or chaotic trajectories can emerge depending on the initial particle separations. 

Chaos is a rarity in flows dominated by viscosity -- Stokes Equations governing the fluid flow are linear and obey kinematic and time reversibility -- however many body interactions can provide a source of nonlinearity in the system \cite{Pine05}. For example, three Stokeslets (the simplest model for a sedimenting particle) can exhibit chaos \cite{Caflisch88, Janosi97}, whereas populations of stresslets (the simplest model for a micro-swimmer) can display coherent large-scale motions \cite{Saintillan12, Lushi13}. Spinning particles (rotors) have been observed to organize in regular arrays and crystals \cite{Grzybowski01,Grzybowski02, Climent07} or exhibit collective directed motion \cite{Grzybowski00}. The regular crystal structuring has been explained by the balance of hydrodynamic repulsion and the magnetic attractive force. 

We analyze here the collective dynamics of purely rotlet particles and show that unlike Stokeslets, the three body interaction does not lead to chaos. The similarities to the 2D point-vortex analogue problem are discussed. In the end we note the implications of pair or triplet interactions in the dynamics of denser rotor suspensions, and outline possible future research directions.

\vspace{0.3in}

{\bf Model:}   
The simplest model for a rotor is a  neutrally buoyant sphere of radius $a$ spinning due to the action of an externally imposed torque $\bT$. In the creeping flow limit where inertia is negligible the generated flow is
\begin{align}
\mathbf{u}(\mathbf{x}) = \frac{1}{8 \pi \mu} \bT \times (\mathbf{x} - \mathbf{x}_i) \frac{a^3}{r^3_i}
\end{align}
which describes a rotlet or {\it couplet} that decays as $1/r^2$ with distance $r = |\mathbf{x} - \mathbf{x}_i |$ from the rotor center (see \cite{KimKarrilla} for the derivation and details). In a suspension, a few well-separated rotors interact with each-other directly and through the fluid: each particle disturbs the fluid flow as it rotates and that in turn affects the other particles. To leading order, a rotor is simply advected by the flow generated by the other rotors
\begin{align}\label{xidot}
\frac{d\mathbf{x}_i}{dt} =  \frac{a^3}{8 \pi \mu} \sum_{j \neq i}  \mathbf{T} \times (\mathbf{x} - \mathbf{x}_j)\frac{1}{ r_{ij}^3}.
\end{align}
where $r_{ij} = |\mathbf{x}_i-\mathbf{x}_j|$ are the separation distances. 

\begin{figure}[htps]
\vspace{-0.15in}
\centering
\includegraphics[width=0.45\textwidth]{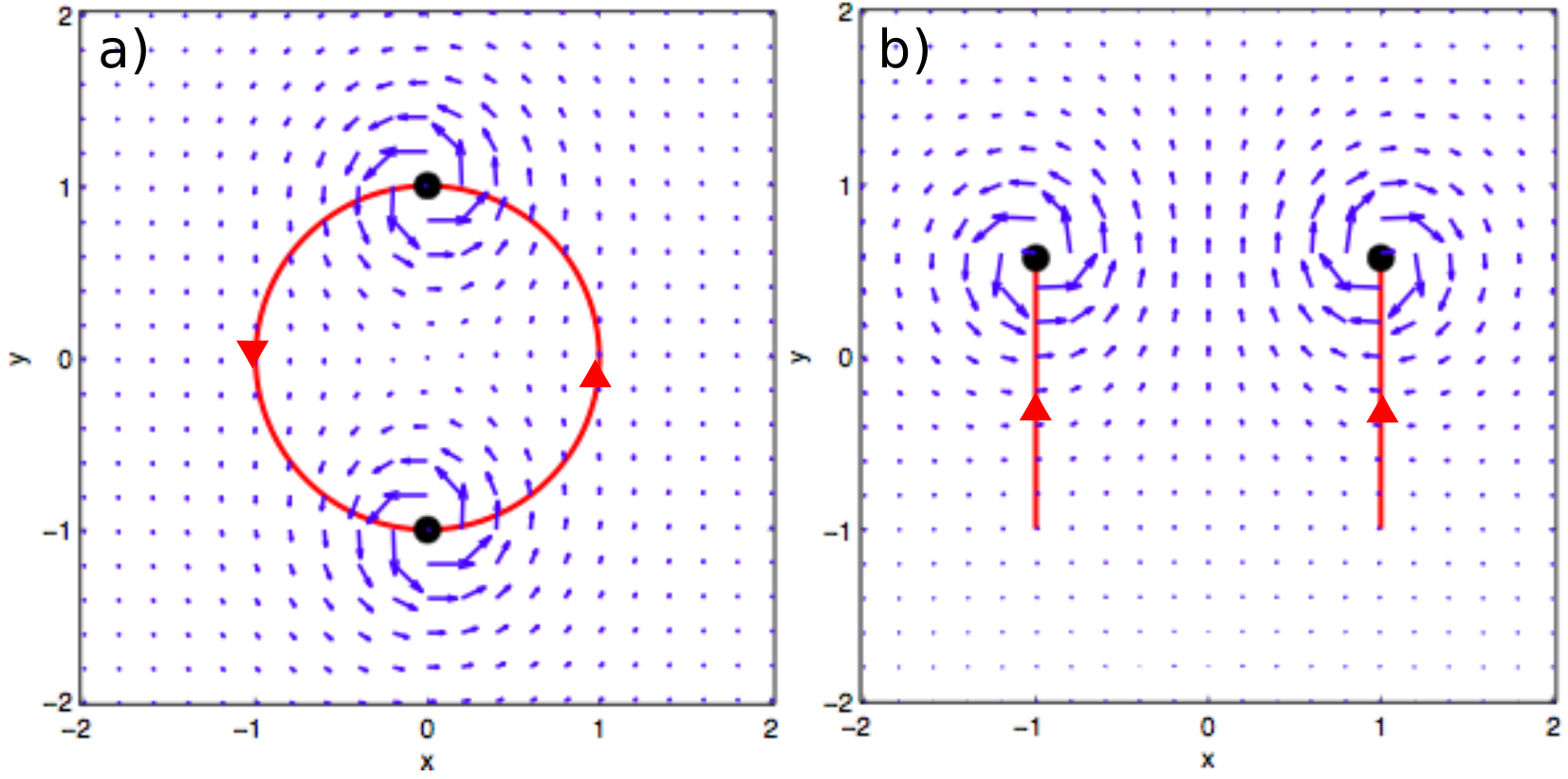}
\vspace{-0.1in}
\caption[width = 0.55\textwidth]{  (Color online) (a) A pair of same-spin rotors circulates around their center of mass. (b) A pair of opposite-spin rotors self-propels in the direction perpendicular to their separation. The fluid flow is shown. Here $|\kappa_i|$=$1$.}
\vspace{-0.15in}
\label{Fig1}
\end{figure}

The dynamics of a pair of rotors with aligned and same magnitude torques that are perpendicular to the rotors' plane of motion is known \cite{Leoni10, Fily12b}. 
Same-spin rotors just orbit around  their centroid at constant speed, see Fig. \ref{Fig1}a, while opposite-spin rotors translate at constant speed in a direction perpendicular to their separation, as illustrated in Fig. \ref{Fig1}b. This form of co-operative self-propulsion between two rotors with different spins has been noted by Leoni and Liverpool \cite{Leoni10} for non-spherical rotors and Fily {\it et. al.} \cite{Fily12b} for purely rotating or rotating-and-translating particles.

Here we show that larger collections of rotating spheres undergo some very surprising dynamics.

{\bf Planar motion of point rotors:} 
Let us first analyze a simpler scenario where all rotors  
lie in the same $x-y$ ($z=0$) plane and all respond to a torque that aligns with the z-axis $\mathbf{T}=T\mathbf{\hat{z}}$. In this case the particle trajectories from Eq. (\ref{xidot}) result in planar motions only ( $d z_i/dt = 0$).
\begin{align}
\frac{d x_i}{dt} &=  \sum_{j \neq i} \frac{ -\kappa_j(y_i-y_j)  }{  \left( (x_i-x_j)^2+  (y_i-y_j)^2 \right) ^{3/2} } \label{xdot}\\
\frac{d y_i}{dt} &=\sum_{j \neq i} \frac{ \kappa_j(x_i-x_j)  }{  \left( (x_i-x_j)^2+  (y_i-y_j)^2 \right) ^{3/2} } \label{ydot}
\end{align}
where   $\kappa_i = Ta^3/8\pi\mu$  can be positive or negative depending on the rotor spin. 
Note that this flow field is azimuthal, i.e. there's no $\mathbf{\hat{z}}$ component. The rotors remain in the $x-y$ plane and the dynamics is essentially 2D. 

The (viscous) rotor dynamics Eqs. (\ref{xdot}-\ref{ydot})
looks similar to the the  2D (inviscid) point-vortex dynamics \cite{Aref_review} except that the power of the separation in the denominator in Eqs. (\ref{xdot}-\ref{ydot}) is $3$ and not $2$.  

Eqs. (\ref{xdot}-\ref{ydot}) for rotors are a Hamiltonian system 
\begin{align}
\kappa_i\frac{d x_i}{dt} = \frac{\partial \mathcal{H}}{\partial y_i}, \quad \kappa_i\frac{d y_i}{dt} = -\frac{\partial \mathcal{H}}{\partial x_i},
\end{align}
with the Hamiltonian
\begin{align}
\mathcal{H} = \kappa_i \kappa_j \sum_{j \neq i} r_{ij}^{-1}, \quad r_{ij} = |\mathbf{x}_i-\mathbf{x}_j| \label{hamilt}
\end{align}
(In 2D point-vortex dynamics it is $\mathcal{H}_V = \kappa_i \kappa_j \sum \log r_{ij}$.)
 The Hamiltonian in Eq. (\ref{hamilt}) is invariant under translation and rotation, and it can be derived by considering that
\begin{align}
&\sum_i \kappa_i x_i = Q=\text{const}, \quad \sum_i \kappa_i y_i = P=\text{const}, \label{sumxi}\\
&\sum_i \kappa_i (x^2_i+ y^2_i) = \mathcal{I}=\text{const}. \label{sumxi2}
\end{align}
This implies that if $\sum_i \kappa_i\neq 0$ the centroid of the point rotlets remains in place and can be chosen as a reference point in the flow. It follows from Eqs. (\ref{sumxi}-\ref{sumxi2}) that 
\begin{align}
\frac{1}{2}\sum_{i\neq j} \kappa_i \kappa_j r^2_{ij} = \left(\sum_i \kappa_i \right) \mathcal{I} - Q^2 - P^2 \label{sumr2}
\end{align}
is a constant of motion that is independent of the coordinates. This a well-known result for point-vortex motion found in most fluid dynamics textbooks. 

Moreover, with a straight-forward calculation, we can show that 
\begin{align}\label{r2dot}
\frac{d}{dt} r_{ij}^2 =  4\sum_{k\neq i \neq j} \kappa_k \mathcal{A}_{ijk}  \left( \frac{1}{r_{ik}^3}- \frac{1}{r_{jk}^3} \right)
\end{align}
where $\mathcal{A}_{ijk}$ is the area of the triangle with sides $r_{ij}, r_{ik}, r_{jk}$ which can be found in terms of them by Heron's formula \cite{ArefGrobli},
 $16\mathcal{A}_{ijk}^2 = \sum_{i \neq j \neq k} (2r^2_{ij}r^2_{ik}-r^4_{ij})$. 
 
Note that the equivalent of Eq. (\ref{r2dot}) for rotors in the analogue 2D point-vortex dynamics involves the term $( r_{ik}^{-2}-r_{jk}^{-2} )$ instead.

Eqs. (\ref{r2dot}) together with the two integrals of motion Eqs. (\ref{hamilt}, \ref{sumr2}) specify the relative motion of $N$ plane-bound rotors in terms of their separations. 
Moreover, the structure of Eq. (\ref{r2dot}) suggests that the three rotor problem is an important block to understanding the dynamics for larger $N>3$ rotors.

\begin{figure}[htps]
\vspace{-0.in}
\centering
\includegraphics[width=0.45\textwidth]{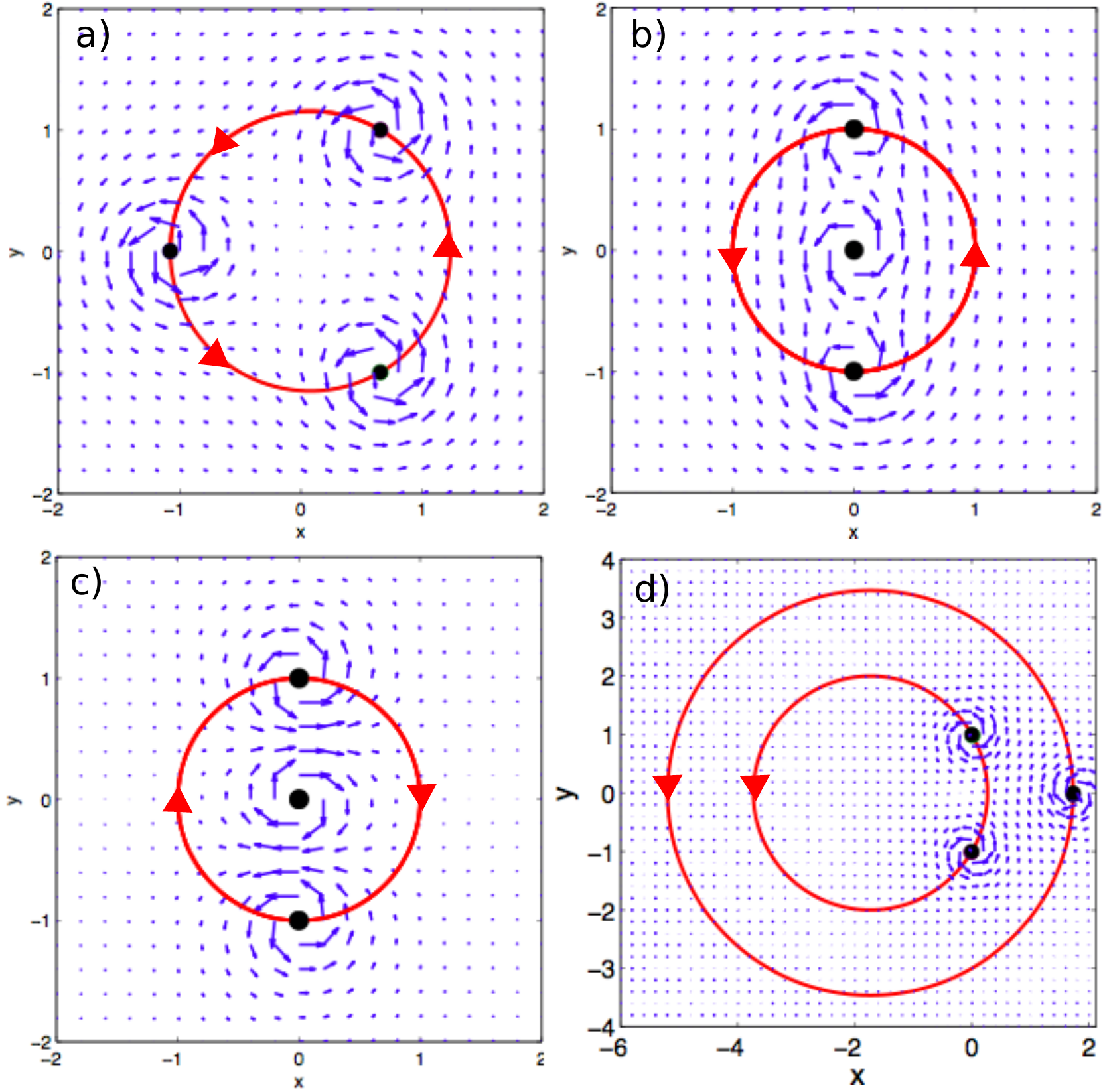}
\vspace{-0.1in}
\caption[width = 0.55\textwidth]{  (Color online) (a) An equilateral triangle configuration of same-spin rotors circulates around their center of mass. Collinear triplet of rotors: the outer ones circulate around the center of mass when the middle rotor has (b) same spin or (c) opposite spin as them.  Note the different orbiting directions. (d) A stable equilateral triangle configuration where the rotor with different spin from the others circulates in the outer trajectory. The fluid flow is shown. In all cases $|\kappa_i|$=$1$.}
\vspace{-0.1in}
\label{Fig2}
\end{figure}

\begin{figure}[htps]
\vspace{-0.in}
\centering
\includegraphics[width=0.45\textwidth]{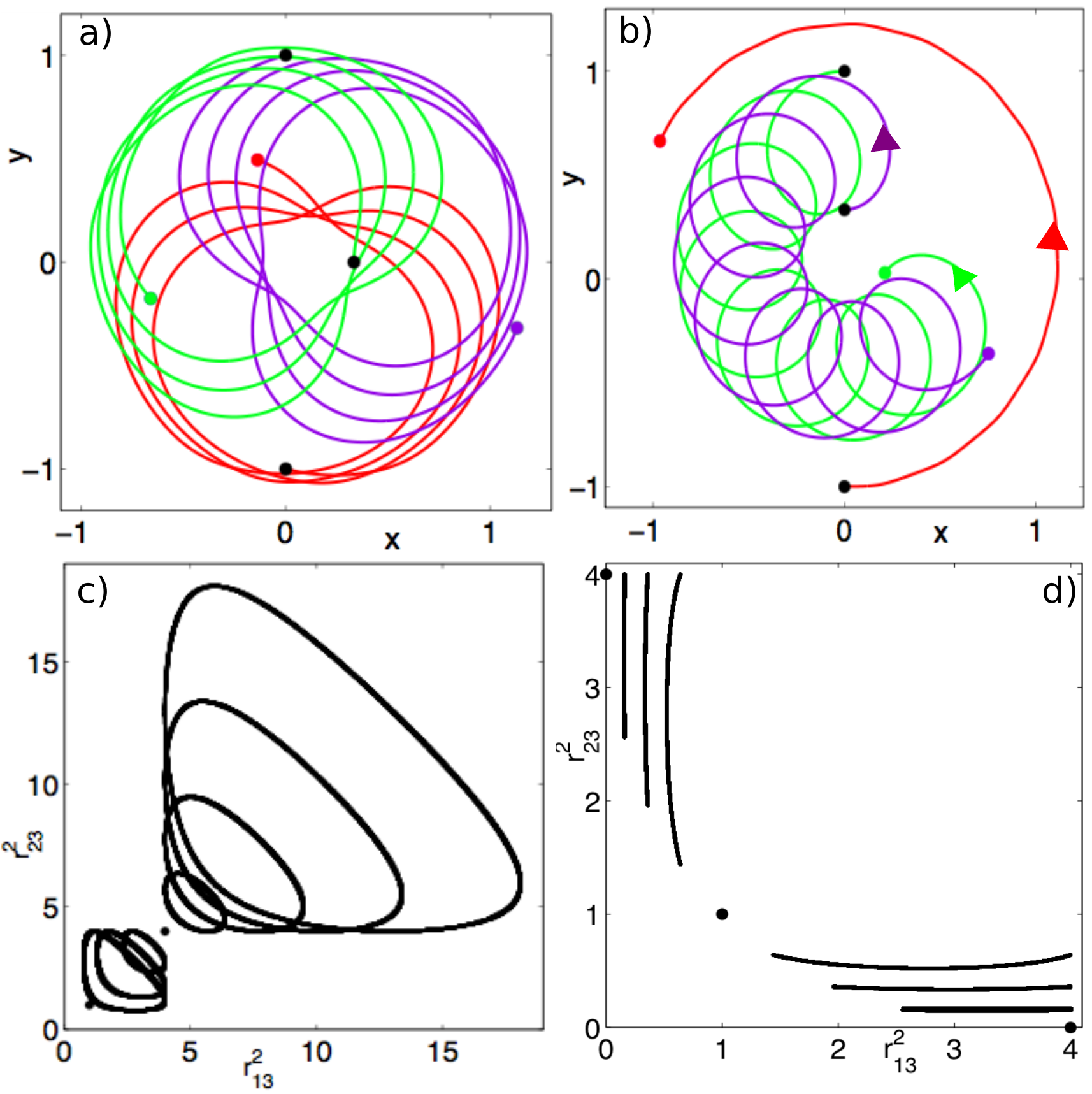}
\vspace{-0.1in}
\caption[width = 0.55\textwidth]{  (Color online) (a) The trajectories of a triplet of same-spin rotors where the initial configuration (shown in black dots) is a horizontal perturbation of the placement of the middle rotor from Fig. \ref{Fig2}ab with $\mathbf{x}_3=(1/4,0,0)$. (b) Trajectories of same-spin rotors when the initial configuration has a vertical perturbation of the placement of the middle rotors from Fig. \ref{Fig2}b with $\mathbf{x}_3=(0,1/4,0)$. (c) The time evolution of the set $(r_{13}^2, r_{23}^2)$ for three-rotor examples where the middle rotor initial placement is a horizontal perturbation shows as in (a) shows closed curves and periodicity. The points $(1,1)$ and $(4,4)$ correspond to the stable crystal configurations in Fig. \ref{Fig2}ab.  (d) The time evolution of the pair $(r_{13}^2, r_{23}^2)$ for examples where the middle rotor initial placement is a vertical perturbation as in (b) also indicates periodicity. The points $(1,1)$, $(4,0)$ and $(0,4)$ show stable crystal configurations, with the last two representing cases where two of the rotors overlap. }
\vspace{-0.1in}
\label{Fig3}
\end{figure}

{\bf Three rotor motion:}
The three-body motion is a classical problem in Physics and Mathematics: e.g. the gravitational mass analogue has been long investigated from Poincar\'{e} himself and is one of the systems known to display chaos. The three-body problem in 2D vortex dynamics however does not display chaos \cite{Aref_review}. Like it, the three rotor problem does not display chaos as it also belongs to a family of integrable systems. While we do not present a formal proof here on the integrability of the three rotor problem, it is possible to construct it following the analogue proof in 2D vortex dynamics \cite{ArefGrobli}. With Eq. (\ref{r2dot}) for the evolution of the three separations, solving for the instantaneous shape of the triangle with the rotors as vertices now only requires elimination of variables and quadrature. The motion can be entirely determined if we have one additional equation in which one or more coordinates and time appear. This is not to say that determining the rotor trajectories is trivial. 

Two special cases of the three-rotor problem are of particular interest: the identical rotors $\kappa_i=\kappa$, $i=1,2,3$, or also the triplet $\kappa_1=\kappa_2=-\kappa_3$. The initial conditions of the problem are very important in the resulting trajectories and dynamics. We investigate this question numerically by integrating Eqs. \ref{xdot}-\ref{ydot} and also making use of the constants of motion Eqs. (\ref{hamilt},\ref{sumr2}) for high accuracy. 

A stable configuration of three same-spin rotors is the equilateral triangle placement: if initially equispaced they remain so and circulate around their center of mass, as seen in Fig. \ref{Fig2}a.  If three rotors are initially collinear with one exactly at the mid-point, then the middle one stays in place while the outer ones circulate around it maintaining collinearity, as seen in Fig. \ref{Fig2}bc, no matter the spin of the middle rotor.  The angular speed of the outer rotors changes however due to the influence of the middle rotor: circulation is faster in all cases compared to the two same-spin rotor case in \ref{Fig1}a . In Fig. \ref{Fig2}ab circulation is counter-clockwise, in Fig. \ref{Fig2}c it is clockwise.

Perturbing the stable configurations of Fig. \ref{Fig2}ab by displacing the middle rotor horizontally (x-direction) or vertically (y-direction) yields an interesting triplet dynamics, as seen in Figs. \ref{Fig3}ab. A horizontal perturbation of the middle (3rd) rotor (hence initial isosceles configuration, yields the triplet dynamics seen in Fig.\ref{Fig2}a where the trajectories are inter-mingled but nonetheless confined within a compact space. A vertical perturbation of the 3rd rotor gives an interesting dynamics as well:  the two nearest rotors pair up and rotate around each-other while also circulating on the whole in a larger motion opposite the other rotor which traces a larger outer trajectory. The trajectories are periodic, and this can be confirmed by looking in Fig. \ref{Fig3}cd at the curves of the set $(r_{13}^2, r_{23}^2)$ in time for a variety of initial conditions such as those in Figs. \ref{Fig3}ab.

\begin{figure}[htps]
\vspace{-0.in}
\centering
\includegraphics[width=0.45\textwidth]{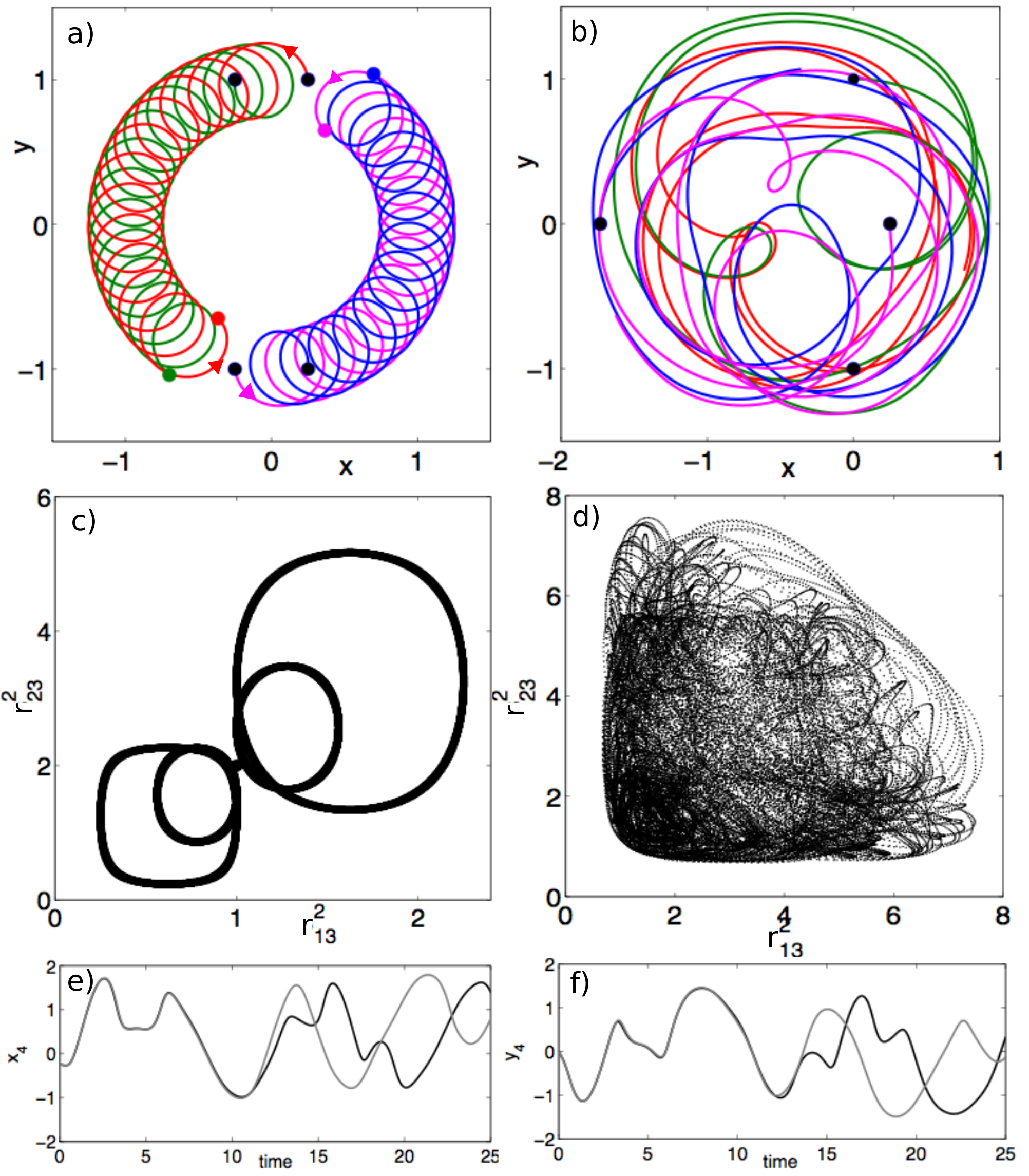}
\vspace{-0.1in}
\caption[width = 0.55\textwidth]{  (Color online) (a) The regular trajectories of four same-spin rotors where the initial configuration (shown in black dots) is a rectangle. (b) Irregular (chaotic) trajectories of four rotors (initial positions shown in black dots). (c) The time evolution of the pairs $(r_{13}^2, r_{23}^2)$ plotted in time for the initially-rectangular configurations as in (a). They are closed curves and indicate periodicity. The point $(1,2)$ corresponds to the stable unit square configuration. (d) The time evolution of the pair $(r_{13}^2, r_{23}^2)$  for the example in (b) suggests chaotic motion. (e,f) $x_4(t)$ and $y_4(t)$ diverge significantly from each other when the initial placement of rotor $\mathbf{x}_4$ from (b) is slightly perturbed by $10^{-3}$ (original trajectories shown in black, their perturbations are shown in gray). The Lyapunov exponent here is positive and, since the trajectories of all the rotors stay in the bounded space seen in (b), this suggests chaotic motion. }
\vspace{-0.in}
\label{Fig4}
\end{figure}

{\bf Four or more rotors:} Just as with the two and three rotor cases, the four rotor problem also can have configurations that stay stable: for example four equal-spin rotors that are initially placed in the vertices of a square will rotate around their centroid (not shown) just as the two or three same-spin rotor analogues. If four same-spin rotors are initially placed in the vertices of a rectangle, we see the dynamics exemplified in Fig. \ref{Fig4}a: the closest rotors pair-up and circulate around each-other while also circulating around the quartet's centroid. The trajectories are periodic, and this can be ascertained by the method proposed by Aref and Pomphrey for four 2D point-vortices \cite{ArefPomphrey80}: we monitor the separations and record the time-dynamics of the pair $(r_{13}^2, r_{23}^2)$ for initial $\dot{r}_{12}>0$ and various initial values of $r_{12}^2=C$; this results into a one-parameter ($C$) family of sections. Since the rotors are identical (if with same spin), any configuration produces $4!$ initial conditions with the same values of the integrals of motion Eqs. (\ref{hamilt}, \ref{sumr2}). Fig. \ref{Fig4}b shows examples of sections and was produced by perturbing in a rectangle configuration. The apparently smooth curves suggest integrability in the vicinity of the stable uniformly rotating square-configuration (aspect ratio unity), and this is consistent with the KAM theory as well \cite{Aref_review}.

By contrast, a non-symmetric initial configuration of the four same-spin rotors yields irregular trajectories seen in an example in Fig. \ref{Fig4}c, which nonetheless stay within a bounded space.  The fourth rotor initial placement in this case is a perturbation of the centered equilateral triangle configuration. Fig. \ref{Fig4}d shows the Poincare section of the $(r_{13}^2, r_{23}^2)$ evolution in time that results from this rotor initial configuration. The section shown contains over $10^5$ points and represents in effects a probability density in the phase space. This kind of random splatter of points, which has been produced with shorter and time-reversible runs as well, suggests chaotic motion, just like in the case of  point-vortices \cite{ArefPomphrey80, ArefPomphrey82}. In Fig. \ref{Fig4}e-f we plot the coordinates of the fourth rotor versus time in for the case in Fig. \ref{Fig4}c and with a very small perturbation of size $10^{-3}$. The trajectories of the two cases quickly de-correlate and diverge from each-other in time, indicating sensibility to initial conditions.

{\bf Effect of the third dimension:} As mentioned before, due to the azimuthal nature of the generated rotlet flow fields that dominate the interactions, the rotors remain in the $x-y$ plane they are initially in. This is also the case when the rotors are initially in different planes $z=constant$: they only move transversely since $dz_i/dt=0$ due to $\mathbf{T}=T\mathbf{\hat{z}}$ (see Eq. (\ref{xidot})). This is obviously a major difference from the 2D point-vortex dynamics. 

For an easy example, a z-direction perturbation placement of another rotor from the centroid of the stable configuration of Fig. \ref{Fig2}abc does not change the overall structure of the ensemble because of the symmetry; only its rotational speed is affected. The three dimensional dynamics of a different configuration can be more difficult to predict analytically, however due to the azimuthal structure of the rotlet flows, the rotor trajectories remain in $z=constant$ planes.

\begin{figure}[htps]
\vspace{-0.1in}
\centering
\includegraphics[width=0.45\textwidth]{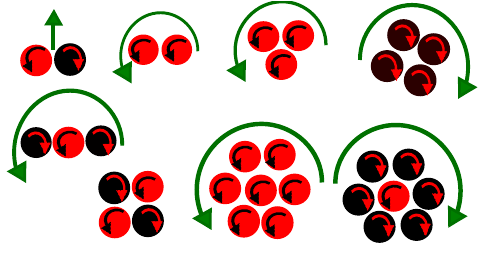}
\vspace{-0.15in}
\caption[width = 0.55\textwidth]{  (Color online) Illustrations of some stable planar crystal rotor structures. Translational or rotational motion of the crystal is indicated with a green arrow when possible, as are the rotor spins (black/red for CW/CCW respectively).}
\vspace{-0.in}
\label{Fig5}
\end{figure}

{\bf Stable rotor crystals:} As we illustrated for 2-3 rotors in Figs. \ref{Fig1}, \ref{Fig2}, for certain initial conditions and separations the rotor configurations remain the same. We can deem them rotor crystals: rotor patterns that move without change of shape or size. It is possible to find crystals for $N>3$ rotors by solving the equations of motions. For 2D point vortices many of such crystal configurations are well-known and well-studied; for example see the reviews \cite{Aref03, NewtonChamoun09}. It is remarkable that the planar crystals of rotors, whose configurations can be solved from Eqs. (\ref{xdot}, \ref{ydot}), are indeed similar to those of vortex crystals, even though the translational or rotational motion of the crystal is not necessarily the same. For completeness, we include some planar rotor crystals as illustrations in Fig. \ref{Fig5}. Unlike the 2D vortex system however, the rotor systems can have crystals that live in three dimensions. For example, it is easy seen that adding rotors of either spin at positions $(0,0,z=constant)$ to the crystals in Figs. \ref{Fig1}, \ref{Fig2} would not destabilize their configurations, though it does affect the crystal translation or rotation. More elaborate 3D crystal structures can be undoubtedly be found analytically or numerically.

{\bf Tracer and passive particles:} Just as in the (2D inviscid) point-vortex case \cite{Neufeld97}, it is possible to study the dynamics of passive particles that are inter-dispersed in a rotor suspension. The dynamics of tracers, or passive particles, though not trivial, is important for understanding mixing of materials at the micro-scale. Some types of active particles, for example micro-swimmers and bacteria, have been explored in simulations \cite{Saintillan12} and experiments \cite{Kim04} as successful agents that can collectively mix a passive field at a faster rate than just by molecular diffusion. Rotors present an exciting possibility to achieve mixing. Analytically studying the dynamics of passive particles in a rotor ensemble is undoubtedly difficult; the analogue problem in 2D point-vortex dynamics, a restricted $N$-body problem, is complicated even for three vortices \cite{Neufeld97}. Numerical studies are however possible.

{\bf Effect of rotor density:} The dynamics of the point-rotors presented thus far rests on the assumptions that the rotors are significantly far from each-other and the dynamics of the flows they generate by rotation can be approximated by the rotlet singularity (see Eq. (\ref{xidot})). If the rotor suspensions are denser (as  they often are in experiments), then the dynamics gets far more complicated. The full hydrodynamics has to be accounted for, namely the higher order singularities in the fluid flow can no longer be neglected as they do modify the particle translations as well as rotations at closer distances. Moreover direct particle collisions can occur and the lubrication flows between the particles cannot be neglected either. Analytically determining the dynamics of even three such rotors is very difficult \cite{KimKarrilla}. Numerical approaches such as the Force Coupling Method \cite{Yeo10a} or Accelerated Stokesian Dynamics \cite{SierouBrady01} however could be successfully employed for this task. Indeed, fluid-mediated interactions can significantly affect the collective rotor behavior \cite{Yeo14}. For example, in our very recent study of binary mixtures of rotors with full hydrodynamical and lubrication interactions, we found that the motion of rotors is surprisingly hindered when the total rotor density in increased \cite{Yeo14}. Certain types of rotor clusters and structures, similar to those illustrated in Fig. \ref{Fig5}, are more prevalent and stable than others, but this depends on the suspension density \cite{Yeo14}. The dynamics of a few rotors, as studied here, could prove illuminating in understanding more complicated structures in more complex scenarios.

{\bf Effect of rotor shape, type and confinement:} In experimental systems the rotor particles are not necessarily spherical, solid, or purely rotators. Any deviation from the spherical shape affects the coupled dynamics. In the far-field approximation non-spherical rotors exhibit similar dynamics to the spherical case, however a secondary dynamics may emerge due to the shape and close-range fluid flows. For example in rotor pairs where each rotor consisted of two beads connected by a thin rod, other than the primary dynamics as in Fig. \ref{Fig1}, a weaker oscillation of each rotor is observed \cite{Leoni10}. Note that in experimental systems the rotating torque does not necessarily remain in the same direction or constant. In fact, the motion of one Quincke rotor in fluid by itself is chaotic \cite{Salipante13} if the full electro-hydrodynamics is accounted for. The rotational motion of a spherical colloid Quincke rotor gets converted into translational motion when they are at a surface, giving rise to the so-called Quincke rollers \cite{Bricard13}. The emergence of collective motion in those confined suspensions is a complex but fascinating phenomenon that is motivating many new studies.

{\bf Discussion and Conclusion:}
We have shown that point rotors in viscous flows can exhibit qualitatively similar behavior to 2D point vortices, e.g., the three rotor motions cannot become chaotic as their dynamics is integrable and the trajectories are periodic. Our work highlights many intriguing analogies and differences between viscous rotors and inviscid vortices. The study suggests many potential studies stemming in such systems, e.g. the periodic or chaotic trajectories of rotors, the structure of stable crystals, the mixing of passive particles or fields. Analytical and computational studies of simple model systems can give important insight into the dynamics observed in more complex scenarios.  


{\bf Acknowledgements:} The authors gratefully acknowledge support from the NSF through nsf-cbet 1437545.


\begin{thebibliography}{26}

\bibitem{Marchetti13} 
M.C. Marchetti, J.F. Joanny, S. Ramaswamy, T.B. Liverpool, J. Prost, M. Rao, and R.A. Simha,
Hydrodynamics of soft active matter.
Reviews of Modern Physics {\bf85}, 1143 (2013).

\bibitem{Menzel15}
A.M.~Menzel
Tuned, driven, and active soft matter.
Physics Reports, {\bf554}, 1-45 (2015).
 
\bibitem{Mallouk09} 
Y.~Wang, S. Fei, Y.-M. Byun, P.E.~Lammert, V.H.~Crespi, A.~Sen and T.E. Mallouk,
Dynamic interactions between fast microscale rotors.
Journal of the American Chemical Society, {\bf 131} (29), 9926-9927 (2009).

\bibitem{Grier97} 
D.G.~Grier, 
Optical tweezers in colloid and interface science.
Current Opinion in Colloid \& Interface Science {\bf2} (3), 264-270 (1997).

\bibitem{Friese98} 
M.E.J.~Friese, T.A.~Nieminen, N.R.~Heckenberg and H.~Rubinsztein-Dunlop, 
Optical alignment and spinning of laser-trapped microscopic particles.
Nature, {\bf 394}, 348-350 (1998).

\bibitem{Grzybowski00} 
B.S. Grzybowski, H.A. Stone, G.M. Whitesides, 
Dynamic self-assembly of magnetized, millimetre-sized objects rotating at a liquidÐair interface.
Nature {\bf405}, 1033-1036 (2000).

\bibitem{Grzybowski02}
B.A. Grzybowski and G.M. Whitesides, 
Dynamic aggregation of chiral spinners.
Science {\bf296}, 718-721 (2002).

\bibitem{Bricard13} 
A. Bricard, J.B. Caussin, N. Desreumaux, O. Dauchot, and D. Bartolo,  
Emergence of macroscopic directed motion in populations of motile colloids.
Nature, {\bf 503}, 95Ð98 (2013).

\bibitem{Leoni10} 
M. Leoni and T.B. Liverpool, 
Dynamics and interactions of active rotors
Europhysics  Letters, {\bf92}, 64004 (2010).

\bibitem{Fily12b}    
Y. Fily, A. Baskaran, and M.C. Marchetti,
Cooperative self-propulsion of active and passive rotors.
Soft Matter {\bf8}, 3002-3009 (2012).

\bibitem{Climent07} 
E. Climent, K. Yeo, M. Maxey and G. Em. Karniadakis, 
Dynamic self-assembly of spinning particles.
Journal of fluids engineering, {\bf129}(4), 379-387 (2007). 

\bibitem{Llopis08} 
I. Llopis and I. Pagonabarraga,
Hydrodynamic regimes of active rotators at fluid interfaces.
The European Physical Journal E {\bf26}, 103-113 (2008).

\bibitem{Yeo10b}  
K. Yeo and M. R. Maxey, 
Rheology and ordering transitions of non-Brownian suspensions in a confined shear flow: Effects of external torques.
Physical Review E {\bf81}, 062501 (2010).

\bibitem{Jibuti12}
L. Jibuti, S. Rafai, P. Peyla,
Suspensions with a tunable effective viscosity: a numerical study.
Journal of Fluid Mechanics {\bf693}, 345-366 (2012).

\bibitem{Fuerthauer13}
S. F\"{u}rthauer, M. Strempel, S.W. Grill, F. J\"{u}licher,
Active chiral processes in thin films.
Physical review letters {\bf110}, 048103 (2013).

\bibitem{Najafi10} 
N. Uchida and R. Golestanian,
Synchronization in a carpet of hydrodynamically coupled rotors with random intrinsic frequency.
 Europhysics  Letters {\bf89} 50011 (2010).

\bibitem{Das13} 
D. Das, D. Saintillan, 
Electrohydrodynamic interaction of spherical particles under Quincke rotation.
Physical Review E, {\bf87},  043014 (2013).

\bibitem{Pine05} 
D.J.~Pine, J.P.~Gollub, J.F.~Brady, A.M.~Leshansky, 
Chaos and threshold for irreversibility in sheared suspensions.
Nature, {\bf 438}, 997-1000 (2005).

\bibitem{Caflisch88}
R.E.~Caflisch, C. Lim, J.H.C.~Luke, and A.S.~Sangani, 
Periodic solutions for three sedimenting spheres.
Physics of Fluids {\bf 31}, 3175 (1988).

\bibitem{Janosi97}
I.M. J\'{a}nosi, T.~T\'{e}l, D.E.~Wolf, and J.A.C.~Gallas, 
Chaotic particle dynamics in viscous flows: The three-particle Stokeslet problem.
Physical Review E, {\bf56} 2858 (1997).

\bibitem{Saintillan12}
D. Saintillan and M.J. Shelley.
Emergence of coherent structures and large-scale flows in motile suspensions.
Journal of the Royal Society Interface {\bf9} (68), 571-585 (2011).

\bibitem{Lushi13}
E. Lushi and C.S. Peskin, 
Modeling and simulation of active suspensions containing large numbers of interacting micro-swimmers.
Computers \& Structures {\bf122}, 239-248 (2013).

\bibitem{Grzybowski01}
B.A.~Grzybowski, X.~Jiang, H.A.~Stone and G.M.~Whitesides, 
Dynamic, self-assembled aggregates of magnetized, millimeter-sized objects rotating at the liquid-air interface.
Physical Review E, {\bf 64}, 011603 (2001).

\bibitem{KimKarrilla}
S. Kim and S.J. Karrilla,
Microhydrodynamics: Principles and Selected Applications.
ButterworthHeinemann, Boston, 1991. 

\bibitem{Aref_review}
H. Aref, 
Integrable, chaotic, and turbulent vortex motion in two-dimensional flows.
Annual Review of Fluid Mechanics {\bf 15}:345-89 (1983).

\bibitem{ArefGrobli}
H.~Aref, N.~Rott, H.~Thomann, 
Gr\"{o}bli's solution of the three-vortex problem.
Annual Review of Fluid Mechanics {\bf 24} 1-20 (1992).

\bibitem{ArefPomphrey80}
H.~Aref and N.~Pomphrey. 
Integrable and Chaotic Motions of Four Vortices.
Physical Review A {\bf78} 297-300 (1980).

\bibitem{ArefPomphrey82} 
H.~Aref and N.~Pomphrey.
Integrable and chaotic motions of four vortices I. The case of identical vortices.
Proceedings of the Royal Society of London A {\bf380} 359-387 (1982).

\bibitem{Aref03} 
H. Aref. P.K. Newton, M.A. Stremler, T. Tokieda and D.L. Vainchtein,
Vortex crystals.
Advances in Applied Mechanics {\bf39} 1-79 (2003).

\bibitem{NewtonChamoun09}
P.K. Newton and G. Chamoun, 
Vortex lattice theory: A particle interaction perspective. 
SIAM Review, Vol 51, Issue 3, 501-542 (2009).

\bibitem{Neufeld97} 
Z. Neufeld and T. T\'{e}l,
The vortex dynamics analogue of the restricted three-body problem: advection in the field of three identical point vortices.
Journal of Physics A: Math. Gen. {\bf30} 2263-2280 (1997).

\bibitem{Yeo10a}
K. Yeo and M.R. Maxey, 
Simulation of concentrated suspensions using the force-coupling method.
Journal of Computational Physics {\bf229}, 2401-2421 (2010).

\bibitem{SierouBrady01}
A. Sierou and J.F. Brady, 
Accelerated Stokesian Dynamics simulations.
Journal of Fluid Mechanics {\bf448}, 115-146 (2004).

\bibitem{Yeo14} 
K. Yeo, E. Lushi and P.M. Vlahovska,
Emergent collective dynamics of hydrodynamically coupled micro-rotors.
arXiv preprint, arXiv:1410.2878 (2014).

\bibitem{Kim04} 
M.J. Kim and K.S. Breuer,
Enhanced diffusion due to motile bacteria.
Physics of Fluids {\bf16} (9) (2004).

\bibitem{Salipante13}  
P.F. Salipante, P.M. Vlahovska,
Electrohydrodynamic rotations of a viscous droplet.
Physical Review E {\bf88},  043003 (2013).

\end{thebibliography}
\end{document}